\documentclass[a4paper,12pt]{article}
\usepackage{graphicx}
\usepackage{mathtools}
\usepackage{amssymb}
\usepackage{amsfonts}
\usepackage[footnotesize]{caption}
\usepackage[font=scriptsize]{subcaption}
\usepackage[usenames,dvipsnames]{color} 
\usepackage{braket}
\usepackage{cite}
\usepackage{hyperref}
\hypersetup{
   colorlinks=true,       
    linkcolor=Blue,          
    citecolor=Plum,        
    filecolor=magenta,      
    urlcolor=YellowOrange           
}
\usepackage{url}
\usepackage{multirow}
\usepackage{relsize}
\usepackage{fullpage}
\usepackage{makecell}
\usepackage{blkarray}
\usepackage{fullpage}
\usepackage{grffile}
\usepackage{float}
\usepackage[labelfont=bf,textfont=md]{caption}
\usepackage{makecell}
\usepackage{tikz}

\usetikzlibrary{shapes,arrows}

\usepackage[latin1]{inputenc}
\usepackage{ae,aecompl}
\usepackage{amsmath,amssymb,mathrsfs}

\providecommand{\bM}{\bar{M}}
\providecommand{\bH}{\bar{H}}
\providecommand{\cp}{\mathsf{CP}}
\providecommand{\dcp}{\delta_{\cp}}

\providecommand{\bM}{\bar{M}}

\providecommand{\om}{\omega}

\providecommand{\mss}[1]{\mbox{\scriptsize $#1$}}
\providecommand{\tp}{{\mss{\mathsf{T}}}}

\DeclareMathOperator{\re}{\mathrm{Re}} 
\DeclareMathOperator{\im}{\mathrm{Im}} 
\DeclareMathOperator{\diag}{\mathrm{diag}} 

\providecommand{\eq}[1]{\begin{equation} #1 \end{equation}}
\providecommand{\eqali}[1]{\begin{equation}\begin{aligned} #1
    \end{aligned}\end{equation}}
    \providecommand{\subeqali}[2][]{\begin{subequations}\label{#1}\begin{align}
#2    \end{align}\end{subequations}}
\providecommand{\mtrx}[1]{\begin{pmatrix} #1 \end{pmatrix}}
\providecommand{\ums}[2][1]{\ml{\tfrac{#1}{#2}}} 

\providecommand{\ml}[1]{\mbox{\large $#1$}}
\providecommand{\xlink}[1]
  {\href{http://arxiv.org/abs/#1}{arXiv:#1}}

\providecommand{\atm}{{\rm a}}
\providecommand{\sol}{{\rm s}}

\usepackage{bbm}
\providecommand{\id}{{\mathbbm{1}}} 

\newcommand{\newc}{\newcommand}
\newc{\be}{\begin{equation}}
\newc{\ee}{\end{equation}}
\newc{\bea}{\begin{eqnarray}}
\newc{\eea}{\end{eqnarray}}
\newc{\simlt}{~\mbox{\smaller\(\lesssim\)}~}
\newc{\simgt}{~\mbox{\smaller\(\gtrsim\)}~}

\begin{document}

\begin{titlepage}
\begin{center}
{\bf\Large
\boldmath{
Mu-tau symmetry and the Littlest Seesaw
}
} \\[12mm]
Stephen~F.~King$^{\star}$%
\footnote{E-mail: \texttt{king@soton.ac.uk}},
Celso~C.~Nishi$^{\dagger}$%
\footnote{E-mail: \texttt{celso.nishi@ufabc.edu.br}},
\\[-2mm]
\end{center}
\vspace*{0.50cm}
\centerline{$^{\star}$ \it
School of Physics and Astronomy, University of Southampton,}
\centerline{\it
SO17 1BJ Southampton, United Kingdom }
\vspace*{0.20cm}
\centerline{$^{\dagger}$ \it
Centro de Matem\'atica, Computa\c c\~ao e Cogni\c c\~ao, }
\centerline{\it
Universidade Federal do ABC - UFABC, }
\centerline{\it
09.210-170,
Santo Andr\'e, SP, Brazil
}

\vspace*{1.20cm}

\begin{abstract}
{\noindent
Motivated by the latest neutrino oscillation 
data which is consistent with maximal atmospheric mixing and maximal leptonic CP violation,
we review various results in $\mu\tau$ symmetry, then 
include several new observations and clarifications, including identifying
a new general form of neutrino mass matrix with $\mu\tau$ symmetry.
We then apply the new results to the neutrino mass matrix associated with the Littlest Seesaw model,
and show that it approximately satisfies the new general form with $\mu\tau$ symmetry,
and that this is responsible for its approximate predictions of 
maximal atmospheric mixing and maximal CP violation in the lepton sector.
}
\end{abstract}
\end{titlepage}

\section{Introduction}

Although neutrino oscillation experiments have provided the first laboratory 
evidence for new physics beyond the Standard Model (BSM) in the form of neutrino mass and mixing
\cite{nobel}, the nature of neutrino mass and lepton flavour mixing remains unknown~\cite{XZbook,King:2013eh}. 
While T2K consistently prefers an almost maximal atmospheric mixing angle~\cite{Abe:2017uxa}, NO$\nu$A originally excluded maximal mixing at $2.6\sigma$ CL~\cite{Adamson:2017qqn}, but the latest analysis with more data is now consistent with maximal mixing \cite{NOvA_new,capozzi,deSalas:2017kay,Esteban:2016qun}.
Furthermore, although the CP violating Dirac phase relevant for neutrino oscillations has not been directly measured,
the global analyses seem to favour somewhat close to maximal values for this phase as well.
The latest neutrino data therefore seems to be consistent with the hypothesis of maximal atmospheric mixing and maximal CP violation in the lepton sector. 
This could either be a coincidence, or may be pointing towards some underlying symmetry or structure underpinning the lepton flavour sector. 

The leading candidate for a theoretical explanation of neutrino mass and mixing remains 
the seesaw mechanism~\cite{Minkowski:1977sc, Yanagida:1979ss, Gell-Mann:1979ss, Glashow:1979ss, Mohapatra:1979ia}. However the seesaw mechanism
involves a large number of free parameters at high energy, and is therefore difficult to test.
One approach to reducing the number of seesaw parameters is to consider the
minimal version involving 
only two right-handed neutrinos (2RHN), first proposed by one of us~\cite{King:1999mb, King:2002nf}.
In such a scheme the lightest neutrino is massless. 
An early simplification~\cite{Frampton:2002qc}, involved two texture zeros in the Dirac neutrino mass matrix
consistent with cosmological leptogenesis~\cite{Fukugita:1986hr,Guo:2003cc, Ibarra:2003up, Mei:2003gn, Guo:2006qa, Antusch:2011nz,
Harigaya:2012bw, Zhang:2015tea}. Although the normal hierarchy (NH) of neutrino masses, 
favoured by current data, is incompatible with the 2RHN model with two texture zeros~\cite{Harigaya:2012bw, Zhang:2015tea}, the one texture zero case originally proposed~\cite{King:1999mb, King:2002nf} remains viable. 

Recently the Littlest Seesaw (LSS) model has been proposed as a particular type of 
2RHN model with one texture zero, which also postulates 
a well defined Yukawa structure with a particular constrained structure
involving just two independent
Yukawa couplings~\cite{King:2013iva, Bjorkeroth:2014vha, King:2015dvf,BAU,King:2016yvg,Ballett:2016yod},
leading to a highly predictive scheme.
Interestingly, the LSS model predicts close to maximal atmospheric mixing and CP violation, as favoured by current data,
a result which can be understood from analytic results.

On the other hand, traditionally, the predictions of maximal atmospheric mixing arise from 
the notion of interchange symmetry between the muon neutrino 
$\nu_{\mu}$ and the tau neutrino $\nu_{\tau}$, namely $\nu_{\mu} \leftrightarrow \nu_{\tau}$,
known as $\mu\tau$ interchange symmetry in the neutrino sector.
When combined with CP symmetry, such a $\mu\tau$ symmetry,
also known as $\mu\tau$ reflection symmetry,
can also lead to maximal CP violation.
For a review of  $\mu\tau$ symmetry see e.g.\cite{mutau:review} and references therein.

In this paper, motivated by the latest neutrino oscillation 
data which is consistent with maximal atmospheric mixing and CP violation,
we give a timely survey of various results in $\mu\tau$ symmetry,
making several new observations and clarifications along the way.
We then apply the new results to the neutrino mass matrix associated with the Littlest Seesaw model,
and show that it approximately satisfies a general form of $\mu\tau$ symmetry,
and that this is responsible for its approximate predictions of 
maximal atmospheric mixing and maximal CP violation in the lepton sector.

The layout of the remainder of this paper is as follows.
In section~\ref{mutau} we introduce and define different types of $\mu\tau$ symmetry as applied to the PMNS matrix $V$,
the neutrino mass matrix $M_\nu$, and its  hermitean square $H_\nu\equiv M_\nu^\dag M_\nu$.
In section~\ref{rephasing} we give basis invariant conditions on $H_\nu$ leading to maximal atmospheric mixing and maximal CP violation.
In section~\ref{sec:general} we present a general form for $M_\nu$ 
with $\mu\tau$ symmetry leading to maximal atmospheric mixing and maximal CP violation.
In section~\ref{sec:mutau.conj} we show how the $\mu\tau$ conjugation operation can be useful 
for relating different neutrino mass matrices which have the general form of $\mu\tau$ symmetry.
In section~\ref{sec:examples} we apply the results to the LSS mass matrix and show why this model
has approximate $\mu\tau$ symmetry. In section~\ref{sec:accidental} we discuss  
accidental implementations of $\mu\tau$ symmetry and give an example.
Finally section~\ref{conclusions} concludes the paper.
The Appendices contain some of the proofs of results in the paper.
Appendix~\ref{proof1} provides a proof that a $\mu\tau$ symmetric $H_\nu$ 
implies and is implied by $\mu\tau$ symmetric PMNS mixing.
Appendix~\ref{sec:cp.basis} makes the connection of the general form of $M_{\nu}$ with $\mu\tau$ symmetry with CP transformations.

\section{Other types of {\large $\mu\tau$} symmetry: $\mu\tau$-U and $\mu\tau$-R}
\label{mutau}

Let us denote by $\mu\tau$ universal ($\mu\tau$-U) mixing the PMNS matrix $V$
characterized by the following two conditions: (i) fully nonvanishing first row, 
\eq{
\label{Vej}
|V_{ej}|\neq 0\,,\quad  j=1,2,3,
}
and (ii) equal moduli for the $\mu$ (second) and $\tau$ (third) rows\,\cite{mutau-r:HS,mutau-r:GL}, 
\eq{
\label{mutau-U}
|V_{\mu j}|=|V_{\tau j}|\,,\quad
j=1,2,3.
}
In other words the modulus of the $\mu\tau$-U PMNS matrix elements have the form
\eq{
\label{Vmutau|}
|V|=\mtrx{|V_{e1}|&|V_{e2}|&|V_{e3}|\cr |V_{\mu 1}|&|V_{\mu 2}|&|V_{\mu 3}|\cr |V_{\mu 1}|&|V_{\mu 2}|&|V_{\mu 3}|}\,.
}
One can show within the standard parametrization that conditions \eqref{Vej} and 
\eqref{mutau-U} are equivalent to having nonzero $\theta_{13}$ together with maximal 
atmospheric angle and Dirac CP phase:%
\footnote{%
Also denoted as cobimaximal mixing in Ref.\,\cite{cobimaximal}.
}
\eq{
\label{maximal.theta}
\theta_{13}\neq 0\,,\quad
\theta_{23}=45^\circ,\quad
\dcp=\pm 90^\circ\,,
}
which are consistent with current data.
The condition \eqref{Vej} ensures the first inequality while \eqref{mutau-U} ensures 
the rest.
In fact, condition \eqref{Vej} implies that both $\theta_{13}$  and $\theta_{12}$ are
nontrivial (different from $0$ or $\pi/2$).
Notice that the Majorana phases in $V$ are not constrained.

Harrison and Scott\,\cite{mutau-r:HS} showed that, allowing rephasing transformations from 
the left and from the right,%
\footnote{%
The following rephasing freedom from the left still survives: $V_{\mu k}\to e^{i\alpha}V_{\mu k}$, $V_{\tau k}\to e^{-i\alpha}V_{\tau k}$.
}
any $\mu\tau$-U PMNS mixing matrix $V$ can be cast in the form 
\eq{
\label{U0}
V_0=\mtrx{|V_{e1}|&|V_{e2}|&|V_{e3}|\cr V_{\mu 1}&V_{\mu 2}&V_{\mu 3}\cr V^*_{\mu 1}&V^*_{\mu 2}&V^*_{\mu 3}}\,.
}
Moreover, when all $|V_{ej}|$ are nonzero, i.e., condition \eqref{Vej} is valid, it is guaranteed that not all of the phases in $V_{\mu i}$ can be removed and $V_0$ is essentially complex.
This fact is consistent with the presence of CP violation in \eqref{maximal.theta}.
The form \eqref{U0} can be easily checked by imposing maximal angle and phase in  
\eqref{maximal.theta} in the standard parametrization and applying appropriate rephasing 
transformations; see Ref.\,\cite{cp.mutau:0} for the explicit form.
In Ref.\,\cite{mutau-r:HS} a different proof was originally supplied and the 
restriction \eqref{Vej} was not imposed.

Instead of characterizing the mixing matrix, it is often more interesting to characterize the neutrino mass matrix $M_\nu$ that is responsible for the mixing in the flavor basis where the $\mu\tau$-U PMNS matrix comes from the diagonalization of the neutrino mass matrix.
As condition \eqref{mutau-U} is insensitive to Majorana phases, it is useful to consider the hermitean square $H_\nu\equiv M_\nu^\dag M_\nu$ of the neutrino mass matrix $M_\nu$ 
for both Majorana or Dirac neutrinos.

We say a hermitean or symmetric $3\times 3$ matrix $A$ is $\mu\tau$-reflection ($\mu\tau$-R) symmetric\,\footnote{%
Also denoted as $\mathsf{CP}^{\mu\tau}$ in Ref.\,\cite{cp.mutau}.
}
if
\eq{
\label{mutau-r:A}
P_{\mu\tau}AP_{\mu\tau}=A^*\,,
}
where 
\eq{
P_{\mu\tau}=\mtrx{1&0&0\cr 0&0&1\cr 0&1&0}\,
}
represents $\mu\tau$ interchange.
According to this definition, the hermitean square mass matrix
$H_\nu=H^{\dagger}_\nu$ is $\mu\tau$-R symmetric~\cite{mutau-r:HS} if it has the form
\eq{
\label{Mnu2:mutau-r:form}
H_\nu=\mtrx{A&D&D^*\cr D^*&B&C^* \cr D&C&B}\,,
}
with $A,B$ real and positive while $C,D$ should have irremovable phases ($\im[C^*D^2]\neq 0$).
It can readily be checked that, if the hermitean square mass matrix $H_\nu$ is $\mu\tau$-R symmetric in the flavour basis
(i.e. has the form in Eq.~\eqref{Mnu2:mutau-r:form}), then this leads to a
$\mu\tau$-U PMNS matrix, with the usual predictions of maximal atmospheric mixing and maximal CP violation.
In fact it can be proved that a
$\mu\tau$-U PMNS matrix implies and is implied by $H_\nu$ being $\mu\tau$-R symmetric in the flavour basis
(see Appendix~\ref{proof1}).

For Majorana neutrinos, the complex symmetric mass matrix $M_\nu$ which leads to a $\mu\tau$-R 
symmetric hermitean square mass matrix $H_\nu$ 
(and hence $\mu\tau$-U PMNS matrix) may take the following special $\mu\tau$-R 
symmetric form\,\cite{mutau-r:GL}\,\footnote{%
This form resulting from a model was first proposed in Ref.\,\cite{babu.ma.valle}.
}
\eq{
\label{Mnu:mutau-r:form}
M_\nu=\mtrx{a&d&d^*\cr d&c&b\cr d^*&b&c^*}\,,
}
with real $a,b$ and $\im[c^*d^2]\neq 0$.
It can readily be checked that the mass matrix of the special $\mu\tau$-R 
symmetric form in \eqref{Mnu:mutau-r:form} leads to a $\mu\tau$-R 
symmetric hermitean square mass matrix $H_\nu$ as in 
\eqref{Mnu2:mutau-r:form} when the hermitean square is taken (and hence a $\mu\tau$-U PMNS matrix).
However it is not necessary for $M_\nu$ to be $\mu\tau$-R 
symmetric, in order to lead to a $\mu\tau$-R 
symmetric hermitean square mass matrix $H_\nu$.%
\footnote{These points were alreay made in Refs.\cite{rodejohann,patel} but here we extend their analysis.}
Unlike Ref.\,\cite{mutau-r:HS}, we shortly show that, while 
 $\mu\tau$-U PMNS mixing is equivalent to having a $\mu\tau$-R symmetric $H_\nu$, it is not 
 equivalent to having a $\mu\tau$-R symmetric $M_\nu$ in the case of Majorana neutrinos.
 In other words, \eqref{Mnu:mutau-r:form} is not the most general form of neutrino mass matrix
 with $\mu\tau$ symmetry.
 But, before giving that, we first discuss the basis invariant
 conditions on $H_\nu$ with {\large $\mu\tau$} symmetry.

\section{Rephasing invariants for $H_\nu$ with {\large $\mu\tau$} symmetry}
\label{rephasing}

We should remark that the discussion in the previous section was based on a phase convention where \eqref{U0} or 
\eqref{Mnu2:mutau-r:form} was valid.
If $\mu\tau$-U mixing follows accidentally (Refs.\,\cite{rodejohann,patel} show one way), we do 
not expect $H_\nu$ to be in the form \eqref{Mnu2:mutau-r:form} as the flavor basis is 
unique only up to rephasing of the $e,\mu,\tau$ flavors.
Therefore, for the task of detecting $\mu\tau$-U mixing using $H_\nu$, 
it is more useful to formulate the 
following three rephasing invariant conditions:
\subeqali[mutau:rephinv]{
\label{mutau:rephinv:1}
\im\big[(H_\nu)_{e\mu}(H_\nu)_{\mu\tau}(H_\nu)_{\tau e}\big]
&\neq 0\,,
\\
\label{mutau:rephinv:2}
|(H_\nu)_{e\mu}|=|(H_\nu)_{e\tau}|
\,,\quad
(H_\nu)_{\mu\mu}&=(H_\nu)_{\tau\tau}\,.
}
See appendix \ref{proof1} for more discussions.
Establishing the equivalence between the conditions in \eqref{mutau:rephinv:2} and
the form \eqref{Mnu2:mutau-r:form}
is straightforward in the basis where $(e\mu)$ and $(e\tau)$ 
entries of $H_\nu $ are real and positive after appropriate rephasing transformations.
In contrast,  the first condition in \eqref{mutau:rephinv:1} is merely the requirement of CP 
violation as, generically, 
\eq{
\label{im.H}
\im\big[(H_\nu )_{e\mu}(H_\nu )_{\mu\tau}(H_\nu )_{\tau e}\big]
=
(m^2_1-m^2_2)(m^2_2-m^2_3)(m^2_3-m^2_1)J\,,
}
where $m_i$ are neutrino mass eigenvalues and $J$ is the usual Jarlskog invariant,
\eq{
J=\im[V_{e1}V^*_{\mu 1}V_{\mu2}V^*_{e2}]
=c_{12}s_{12}c_{13}^2s_{13}c_{23}s_{23}\sin(\delta)\,.
}
Note that \eqref{im.H} is $\im[C^*D^2]$ in the notation of \eqref{Mnu2:mutau-r:form}
and the sign of \eqref{im.H} is given by the sign of $J$ for physical cases.

The conditions in \eqref{mutau:rephinv}
do not seem to be clearly written in the literature. For example, 
the review \cite{mutau:review} only mentions the first condition in 
\eqref{mutau:rephinv:2}.
Generically, only one of the conditions in \eqref{mutau:rephinv:2} is not enough and 
simple numerical examples quickly show that deviation from any of the two conditions 
spoils \eqref{maximal.theta}.
Another important aspect is that these conditions can be simplified if specific phase 
conventions are adopted.

Specifically for Majorana neutrinos, since a $\mu\tau$-U mixing matrix in the form 
\eqref{U0} diagonalizes $H_\nu$, it can still diagonalize $M_\nu$ as
\eq{
\label{diag:mutau:Hnu}
V_0^\tp M_\nu V_0=\diag(m_je^{i\alpha_j})\,,
}
where two of the differences of $\alpha_j$ ---the Majorana phases--- 
are arbitrary and in principle nontrivial. One of $\alpha_j$ can be eliminated 
by a global rephasing.

We can now describe the stronger case of having a $\mu\tau$-R symmetric $M_\nu$ in the 
form \eqref{Mnu:mutau-r:form} for Majorana neutrinos.
In this case, apart from the predictions from $\mu\tau$-U mixing in 
\eqref{maximal.theta}, $\mu\tau$-R symmetry leads to the additional consequence that all 
the Majorana phases are trivial\,\cite{mutau-r:GL}.
The form \eqref{Mnu:mutau-r:form} is more interesting for model building because it can be 
enforced by $\mu\tau$-reflection symmetry at the level of fields as
\eq{
\label{mutau-r:nuL}
\nu_e\to \nu_e^{cp}
\,,\quad
\nu_\mu\to \nu_\tau^{cp}
\,,\quad
\nu_\tau\to \nu_\mu^{cp}
\,,
}
where $\mu\tau$ interchange symmetry is combined with the standard CP transformation 
denoted by $cp$.

The conclusion that Majorana phases are trivial can be drawn as follows.
The diagonalizing matrix $V_0$ in \eqref{diag:mutau:Hnu} has the form \eqref{U0} which exhibits the special property that complex conjugation has the same effect as the interchange of the second ($\mu$) and third ($\tau$) rows.
Then, the additional property that $M_\nu$ itself is $\mu\tau$-R symmetric as in \eqref{mutau-r:A} tells us that 
\eq{
V_0^\tp M_\nu V_0=(V_0^\tp M_\nu V_0)^*=\diag(m_i')\,,
}
implying real $m'_i=\pm m_i$.
Hence Majorana phases are trivial and CP parities can be defined depending on the signs 
of $m'_i$. See Ref.\,\cite{cp.mutau} for generic implications of $\mu\tau$-R symmetry on neutrinoless double beta decay and leptogenesis.

We are now naturally led back to the question: What is the most general form of 
neutrino mass matrix $M_\nu$ which 
has $\mu\tau$-R symmetric $H_\nu$ but may not itself be $\mu\tau$-R symmetric?
This question has been only partially answered in the literature and we devote
Sec.\,\ref{sec:general} to this question.

\section{General form of $M_\nu$ with {\large $\mu\tau$} symmetry}
\label{sec:general}

We characterize here all the Majorana neutrino mass matrices $M_\nu$ whose hermitean square is $\mu\tau$-R symmetric as in \eqref{Mnu2:mutau-r:form}.
We stress that $M_\nu$ itself does not need to be $\mu\tau$-R symmetric, i.e., it does not 
have to take the form \eqref{Mnu:mutau-r:form}.

The main result is that any $M_\nu$ can be always decomposed into two $\mu\tau$-R symmetric matrices $A,B$, not both zero, as
\eq{
\label{mutau-r:decomp}
M_\nu=A+iB\,,
}
and $M_\nu$ has $\mu\tau$-U mixing if and only if
\eq{
\label{mod.comm}
A^*B=B^*A\,,
}
provided that \eqref{mutau:rephinv:1} ensures one irremovable phase.
Of course, in this case, $H_\nu=M_\nu^\dag M_\nu$ is both $\mu\tau$-R and $\mu\tau$-U symmetric.

The first decomposition \eqref{mutau-r:decomp} can be always performed, irrespective 
of any symmetry, by
\eq{
\label{A,B}
A=\frac{1}{2}(M_\nu+\widetilde{M_\nu})\,,
\quad
B=\frac{1}{2i}(M_\nu-\widetilde{M_\nu})\,,
}
where we have defined the $\mu\tau$ conjugation operation as
\eq{
\label{mutau.conj}
M\to \widetilde{M}\equiv P_{\mu\tau}M^* P_{\mu\tau}\,,
}
which we denote by a tilde.
By construction both components are $\mu\tau$-R symmetric:
$\widetilde{A}=A$ and $\widetilde{B}=B$.
Note, however, that $iB$ is not $\mu\tau$-R symmetric and it is frequently 
considered as a quantifier of the breaking of $\mu\tau$-R (see review in \cite{mutau:review}). 
Here, we clarify that it is only the component in $A$ or $B$ that does not satisfy \eqref{mod.comm} that leads to the breaking of \eqref{mutau-U}.
So for $A\neq 0$ it is possible to have large breaking of $\mu\tau$-R symmetry in the form of large $B$ but still preserving $\mu\tau$-U mixing.
For example, a large breaking of the $\mu\tau$-R symmetric $M_\nu$ in the form of \eqref{mutau-r:decomp},
\eq{
\label{ix}
M_\nu=\mtrx{a&d&d^*\cr d&c&b\cr d^*&b&c^*}
+i\mtrx{x&0&0\cr 0&0&x\cr 0&x&0}
\,
}
where $x$ is real and not small in general,
leads to $\mu\tau$-U mixing.\footnote{%
Other forms in this direction can be found in \cite{yasue} but the general form was not shown.
}
It follows from the simple fact that the second matrix is proportional to $P_{\mu\tau}$.
We will see another example of a mass matrix with $\mu\tau$-U mixing without $\mu\tau$-R
later in \eqref{Mnu:sp} for which $|(M_{\nu})_{e\tau}|$ is three 
times larger than $|(M_{\nu})_{e\mu}|$.

The proof of condition \eqref{mod.comm} is straightforward.\footnote{%
We thank the anonymous referee for pointing to this route.}
Using the decomposition \eqref{mutau-r:decomp} with \eqref{A,B}, we can write
\eq{
H_\nu=A^*A+B^*B+i(A^*B-B^*A)\,.
}
Since $A,B$ are $\mu\tau$-R symmetric by construction, it is clear that
\eq{
\widetilde{H_\nu}=A^*A+B^*B-i(A^*B-B^*A)\,.
}
Then $A^*B=B^*A$ should hold if $H_\nu$ is $\mu\tau$-R symmetric. The converse is obvious from the expressions above.
One point to note is that CP violation in \eqref{mutau:rephinv:1} should be ensured separately.
For example, $M_\nu=e^{i\alpha}M_0$, where $M_0$ is real and symmetric by 
$\mu\tau$-interchange, has components that satisfy \eqref{mod.comm} but has trivial 
Dirac CP phase.

We should also remark that the decomposition \eqref{mutau-r:decomp} is not rephasing invariant: components $A$ and $B$ get mixed up when the decomposition is performed after some rephasing transformation.
And then condition \eqref{mod.comm} may be satisfied in one basis but not in another connected by rephasing.
We can analyze the different situations by decomposing a general rephasing transformation as the product of the following transformations on the $(e,\mu,\tau)$ flavors:
\eq{
\label{rephasings}
(e^{i\alpha},e^{i\alpha},e^{i\alpha}),~~
(1,e^{i\alpha},e^{-i\alpha}),~~
(1,e^{i\alpha},e^{i\alpha})\,.
}
The first transformation induces a \textit{global} rephasing of $M_\nu$ and $H_\nu$ is invariant.
In this case, $A$ and $B$ get mixed but once \eqref{mod.comm} is satisfied, it is satisfied after any global rephasing with subsequent decomposition.
The second transformation preserves the form of $\mu\tau$-R symmetry \eqref{mutau-r:nuL}
and thus preserves the decomposition \eqref{mutau-r:decomp}: $A,B$ are transformed separately preserving the $\mu\tau$-R symmetric form.
The third transformation in \eqref{rephasings} do not preserve the decomposition \eqref{mutau-r:decomp} and rephasing generically transforms a pair $(A,B)$ that satisfies condition \eqref{mod.comm} into a pair that does not.
We can use the first two transformatios in \eqref{rephasings} to set a convention for $M_\nu$. One possibility is to use global rephasing to choose the $(ee)$ entry real and positive\,\footnote{%
In case that entry is zero, we could choose the $(\mu\tau)$ entry.
See Refs.\,\cite{cp.mutau:0,cp.mutau:0i} for explicit models exhibiting $\mu\tau$-R symmetry
and one texture-zero in the neutrino mass matrix or its inverse.
}
and the second transformation to make the phases of $(e\mu)$ and $(e\tau)$ entries equal.
Another possibility is to use global rephasing so that the phase on $m_1$ in \eqref{diag:mutau:Hnu} is removed.

Let us check the number of parameters using the convention mentioned above.
If we use the phase convention where the $(ee)$ entry of $M_\nu$ is real, we would have $B_{ee}=0$. We can still eliminate one phase in $A$ so that it has 5 parameters, as in \eqref{Mnu:mutau-r:form} with real $d$.
The component $B$ in this basis also has 5 parameters because there is no more rephasing freedom left.
We end up with $5+5-3=7$ parameters from $A,B$ minus the number of conditions from \eqref{mod.comm}.
This number concides with the three masses ($m_i$), two angles 
($\theta_{13},\theta_{12}$), and two independent Majorana phases that appear in 
\eqref{diag:mutau:Hnu}.

We can confirm the number of conditions by defining $C$ by
\eq{
iC=AB^*-BA^*\,.
}
The matrix $C$ is then hermitean but $\mu\tau$-R antisymmetric, i.e., $\widetilde{C}=-C$.
Thus we can write
\eq{
C=\mtrx{0&y&-y^*\cr y^*&x&0 \cr -y&0&-x}\,,
}
with real $x$. Therefore, $C=0$ imposes three conditions.
Explicit calculation leads to $2x=(H_\nu)_{\tau\tau}-(H_\nu)_{\mu\mu}$
and $2y=(H_\nu)_{\tau e}-(H_\nu)_{e\mu}$.
So $H_\nu$ is $\mu\tau$-R symmetric if they vanish.

It is important to stress that our discussions are valid in a basis where $\mu\tau$-R symmetry of $H_\nu$ is manifest.
When it is not known a priori that $\mu\tau$-U mixing holds, one should resort to \eqref{mutau:rephinv} for the truly rephasing invariant criterion.
If one detects that the conditions in \eqref{mutau:rephinv} are satisfied but $H_\nu$ is not explicitly in the form \eqref{Mnu2:mutau-r:form}, rephasing transformations should be applied to get to that form and then $M_\nu$ in the same basis should be taken.
It is in this basis that condition \eqref{mod.comm} holds.

\section{Neutrino mass matrices related by $\mu\tau$ conjugation}
\label{sec:mutau.conj}

Here we show that the $\mu\tau$ conjugation operation defined in \eqref{mutau.conj}, when applied to neutrino mass matrices,
\eq{
\label{mutau.conj2}
M_{\nu}\to \widetilde{M_{\nu}}\equiv \mtrx{1&0&0\cr 0&0&1\cr 0&1&0}M_{\nu}^* \mtrx{1&0&0\cr 0&0&1\cr 0&1&0}\,,
}
can be useful 
for relating the predictions from the different neutrino mass matrices $M_{\nu}$ and $\widetilde{M_{\nu}}$
even when they do not
exhibit $\mu\tau$-U mixing or $\mu\tau$-R symmetry.

Generically, even without $\mu\tau$-U mixing, we can always adopt the phase convention in \eqref{U0} for the PMNS matrix, without Majorana phases, where the first row is real and positive.
Without $\mu\tau$-U mixing, the third row does not need to have the same modulus as 
the second.
Assuming Majorana neutrinos, the diagonalization \eqref{diag:mutau:Hnu} is still valid.
Taking the complex conjugate of such an expression, we conclude that the $\mu\tau$ conjugate of $M_\nu$ is diagonalized as
\eq{
\label{M-tilde:U}
\widetilde{V_0}^\tp\widetilde{M_\nu}\widetilde{V_0}=\diag(m_je^{-i\alpha_j})\,,
}
with Majorana phases flipping sign and 
the $\mu\tau$ conjugate {\em of the mixing matrix} defined as
\eq{
\widetilde{V_0}\equiv P_{\mu\tau}V_0^*\,.
}
For Dirac neutrinos, the conclusion is the same, without Majorana phases.

With $\mu\tau$-U mixing, $\mu\tau$ conjugation is trivial: $\widetilde{V_0}=V_0$.
For generic mixing, the $\mu\tau$ conjugation operation
\eq{
\label{U0:conj}
V_0\to \widetilde{V_0}\equiv  \mtrx{1&0&0\cr 0&0&1\cr 0&1&0}V_0^*,
}
keeps the first row and the Jarlskog invariant unchanged whereas the values for $|(V_0)_{\tau 2}|$ and $|(V_0)_{\tau 3}|$ are interchanged.
Comparing to the standard parametrization, this means that $\theta_{13},\theta_{12}$,
remain unchanged whereas
\eq{
\label{octant}
\theta_{23}\to \pi/4-\theta_{23}\,
}
is mapped to its complement in the opposite octant.
Since the combination that enters the Jarlskog invariant, $\sin\theta_{23}\cos\theta_{23}$, remains the same, the value of $\sin\delta$ is unchanged but $\cos\delta$ changes sign.
Therefore, the Dirac CP phase transforms as
\eq{
\label{cp}
e^{i\delta}\to -e^{-i\delta}\,.
}
Since present neutrino oscillation data is favored by values close to the $\mu\tau$-U mixing, then $\mu\tau$ conjugate mass matrices can be equally appropriate to describe data.
Note that $\mu\tau$ conjugation contrasts with plain complex conjugation on the 
mass matrix and mixing matrix which maintains all mixing angles the same but flips 
all CP phases.

The sign flip of $\cos\delta$ can be confirmed from
\eq{
\label{Vmu-tau:2}
|(V_0)_{\mu 2}|^2-|(V_0)_{\tau 2}|^2=
(c^2_{12}-s^2_{12}s^2_{13})(c^2_{23}-s^2_{23})
-4\cos(\delta)\, s_{12}c_{12}s_{23}c_{23}s_{13}
\,,
}
in the standard parametrization.
As the lefthand side and the first term of the righthand side flip sign under the conjugation \eqref{U0:conj}, $\cos(\delta)$ has to flip sign because its coefficient is invariant.

In summary, two mass matrices $M_\nu,\widetilde{M_\nu}$ related by $\mu\tau$ conjugation in \eqref{mutau.conj2} lead to identical
predictions for $\theta_{13},\theta_{12}$, with $\theta_{23}$ changing octant as in \eqref{octant}, while $\sin \delta$ is unchanged but
$\cos \delta$ changes sign corresponding to the phase transformation in \eqref{cp}. 

Furthermore, $M_\nu,\widetilde{M_\nu}$ related by $\mu\tau$ conjugation will predict opposite Majorana phases,
even in the special case that they are diagonalizable by the same $\mu\tau$-U mixing matrix with maximal atmospheric mixing and 
maximal Dirac CP violation.


\section{Littlest seesaw and approximate {\large $\mu\tau$} symmetry}
\label{sec:examples}

We are now ready to apply the foregoing results to a well known minimal example in the 
literature.
The Littlest Seesaw (LSS) neutrino mass matrix proposed in Ref.\,\cite{King:2016yvg} is a
highly predictive two-parameter neutrino mass matrix given by
\eqali{
\label{Mnu:lss.2}
M_\nu&=m_\atm\mtrx{0&0&0\cr 0&1&1\cr 0&1&1}+\omega m_\sol\mtrx{1&1&3\cr 1&1&3\cr 3&3&9}\,,
}
where $m_{\atm}$ and $m_\sol$ are real and positive while $\om=e^{i2\pi/3}$.
\footnote{The phase $\omega=e^{i2\pi/3}$ is denoted as $e^{i\eta}$ in \cite{King:2016yvg}
and here we define the neutrino mass matrix elements as $(M_\nu)_{ij} \nu_{iL}\nu_{jL}$ in the flavor basis in contrast to $(M_\nu)_{ij} \nu^c_{iL}\nu^c_{jL}$ used in \cite{King:2016yvg}.}
The form \eqref{Mnu:lss.2} leads to TM1 mixing which predicts $s^2_{12}c^2_{13}=\ums{3}(1-3s^2_{13})$\,\cite{TM}.
This mass matrix is closely related to the original LSS mass matrix proposed in \cite{King:2015dvf}:
\eq{
\label{Mnu:lss}
\widetilde{M_{\nu}}=m_\atm\mtrx{0&0&0\cr 0&1&1\cr 0&1&1}+\omega^2 m_\sol\mtrx{1&3&1\cr 3&9&3\cr1&3&1}\,,
}
with real and positive $m_\atm$ and $m_\sol$.
In fact, they are related by $\mu\tau$ conjugation in \eqref{mutau.conj2}.
Since the neutrino parameters from one case can be easily extracted from the other, 
we can focus on \eqref{Mnu:lss.2}.

The best-fit values for $m_{\atm,\sol}$ in \eqref{Mnu:lss.2} are\,\cite{King:2016yvg}
\eq{
m_\atm=26.57\,\text{meV}\,,\quad
m_\sol=2.684\,\text{meV}\,.
}
These values result in $m_1=0, m_2=8.59\,\text{meV}\,,m_3=49.8\,\text{meV}$
and
\eq{
\label{best-fit}
\theta_{23}=44.2^\circ\,,\quad
\dcp=-93.3^\circ\,.
}
The rest of the parameters are $\theta_{12}=34.3^\circ,\theta_{13}=8.67^\circ$.
From $\mu\tau$ conjugation, all the predictions for \eqref{Mnu:lss} are the same as
for \eqref{Mnu:lss.2}, except that $\theta_{23}$ and $\dcp$ get complementary values with respect to $45^\circ$ and $-90^\circ$, respectively.

We can immediately see that the values in \eqref{best-fit} are close to the $\mu\tau$-U 
mixing predictions in \eqref{maximal.theta}.
In fact, one can check that an exact $\mu\tau$-U mixing is obtained if $m_{\atm,\sol}$ 
satisfy the special ratio
\eq{
\label{eps=}
\frac{m_\atm}{m_\sol}=11\,,
}
as can be checked explicitly using the 
analytic formulas in Refs.\,\cite{King:2015dvf,King:2016yvg}.
Inserting this ratio of masses, the neutrino mass matrix in \eqref{Mnu:lss.2} becomes, after multiplying by an overall physically irrelevant
phase of $\omega^2$,
\eq{
\label{Mnu:sp}
M_\nu=
m_\sol
\left(
\begin{array}{ccc}
 1 & 1 & 3 \\
 1 & 1+11 \om^2 & 3+\om^2 \\
 3 & 3+11\om^2 & 9+11\om^2 \\
\end{array}
\right)\,.
} 
Clearly, there is no $\mu\tau$-R symmetry on $M_\nu$, i.e., it does not have the form in 
 \eqref{Mnu:mutau-r:form}.
However, by comparing to \eqref{Mnu2:mutau-r:form}, it is easy to check that its hermitean square,
\eq{
\label{Mnu2:sp}
H_\nu=
M_\nu^\dag M_\nu=
11\,|m_\sol|^2
\left(
\begin{array}{ccc}
 1 & -1-2 i \sqrt{3} & 1-2 i \sqrt{3} \\
 -1+2 i \sqrt{3} & 19 & 17+4 i \sqrt{3} \\
 1+2 i \sqrt{3} & 17-4 i \sqrt{3} & 19 \\
\end{array}
\right)\,,
}
does satisfy $\mu\tau$-R symmetry after we flip the sign of the second row and 
column. Thus we conclude that the LSS mass matrix obeys 
$\mu\tau$-U PMNS mixing in the limit of Eq.\,\eqref{eps=}.
Since the best fit parameters of the LSS model are close to Eq.\,\eqref{eps=} then we can understand why 
its predictions for the atmospheric angle and CP phase are both close to maximal.
However, since the LSS mass matrix has only two input parameters, which fixes all neutrino masses and 
PMNS mixing parameters, there are other predictions including the reactor angle, the solar angle,
the absolute neutrino masses and the Majorana phase, which $\mu\tau$ symmetry by itself does not address.

Obviously, the rephasing invariant conditions in \eqref{mutau:rephinv:2} are also satisfied.
To check that $H_\nu $ is essentially complex, we can use \eqref{mutau:rephinv:1} and obtain
\eq{
\im\big[(H_\nu )_{e\mu}(H_\nu )_{\mu\tau}(H_\nu )_{\tau e}\big]
=-11^3|m_\sol|^6\times 24\sqrt{3}\neq 0\,.
}
Since it is negative, the ambiguity in the sign of $\delta$ in \eqref{maximal.theta}
is now removed and we have $\delta=-\pi/2$ in this case.
Alternatively, we could use rephasing with opposite phases for $\mu$ and $\tau$ on 
\eqref{Mnu2:sp} to eliminate the arguments of the entries $(e\mu)$ and $(e\tau)$ so that
\eq{
H_\nu \to
11|m_\sol|^2
\mtrx{1 & \sqrt{13} & \sqrt{13}
\cr
\sqrt{13} & 19 & \sqrt{337}e^{-i\gamma}
\cr
\sqrt{13} & \sqrt{337}e^{i\gamma} & 19
}
\,,
}
where $\gamma
=\arctan(\frac{24\sqrt{3}}{235})\approx 0.175$.
As expected from the absence of $\mu\tau$-R symmetry on $M_\nu$, one Majorana phase is nontrivial as
\eq{
P=\diag(1,1,e^{-i\beta/2})\,,
}
where $\beta\approx 1.8$ and we use the convention where $V=V_0P$,
with $V_0$ having the first row real and positive.

Since \eqref{Mnu2:sp} is not explicitly in the $\mu\tau$-R symmetric form 
\eqref{Mnu2:mutau-r:form}, we flip the sign of the second row and column of both $H_\nu$ 
and $M_\nu$ in \eqref{Mnu:sp}.
We obtain for $M_\nu/m_\sol$,
\subeqali[M:sp]{
M_\nu/m_\sol&=
\left(
\begin{array}{ccc}
 1 & -1 & 3 \\
 -1 & 1+11 \om^2 & -3-11\om^2 \\
 3 & -3-11\om^2 & 9+11\om^2 \\
\end{array}
\right)\,,
\\
\label{LSS:mutau:M}
&=\mtrx{1&-1&3\cr -1&1&-3\cr 3&-3&9}
+11\om^2 \mtrx{0&0&0\cr 0&1&-1\cr 0&-1&1}
\,,
}
where we automatically have real $(ee)$ entry. 
For illustration, we can apply the decomposition \eqref{mutau-r:decomp}, 
$M_\nu/m_\sol=A+iB$, where 
in this case:
\eqali{
A&=\left(
\begin{array}{ccc}
 1 & 1 & 1 \\
 1 & -\frac{1}{2} & \frac{5}{2} \\
 1 & \frac{5}{2} & -\frac{1}{2} \\
\end{array}
\right)\,,
\cr
B&=\left(
\begin{array}{ccc}
 0 & 2 i & -2 i \\
 2 i & 4 i-\frac{11 \sqrt{3}}{2} & \frac{11 \sqrt{3}}{2} \\
 -2 i & \frac{11 \sqrt{3}}{2} & -4 i-\frac{11 \sqrt{3}}{2} \\
\end{array}
\right)
\,.
\label{AB}
}
Both $A$ and $B$ are $\mu\tau$-R symmetric as in \eqref{Mnu:mutau-r:form} but the sum $A+iB$ is not (since $iB$ is not).
One can check that these two matrices satisfy the modified commutation relation in 
\eqref{mod.comm}.

We conclude that the neutrino mass matrix in \eqref{Mnu:sp}, or its equivalent form \eqref{M:sp}, does indeed satisfy
the general form of $\mu\tau$-R symmetry $M_\nu/m_\sol=A+iB$ where $A,B$ are given in \eqref{AB}.
One may call this the $\mu\tau$ symmetric Littlest Seesaw ($\mu\tau$-LSS).

The matrix \eqref{M:sp} completely determines the mixing matrix which has first row given by
\eq{
(|V_{ei}|^2)=\Big(\frac{2}{3},
\frac{1}{6}+\frac{11}{18 \sqrt{17}},\frac{1}{6}-\frac{11}{18 \sqrt{17}}\Big)
\approx
(0.667,0.315,0.0185)\,.
}
Hence we obtain $\theta_{13}=7.807^\circ,\theta_{12}=34.5^\circ$.
Although $\theta_{12}$ is nicely consistent with global fit data, the value of $\theta_{13}$ is slightly lower than what is allowed within $3\sigma$\,\cite{Esteban:2016qun}.

We can also calculate the eigenvalues,
\eqali{
m_1^2&=0,
\cr
m_2^2&=11m_\sol^2\times\ums[3]{2}(13 - 3\sqrt{17})
\approx (3.226\times m_\sol)^2,
\cr
m_3^2&=11m_\sol^2\times	\ums[3]{2}(13 + 3\sqrt{17})
\approx (20.46\times m_\sol)^2.
}
It is also not possible to fit $m_\sol$ so that we can simultaneously accommodate $\Delta m^2_{21}$ and $\Delta m^2_{31}$ within 3$\sigma$.
We therefore conclude that the $\mu\tau$-LSS neutrino mass matrix cannot give correct predictions in the exact 
$\mu\tau$ symmetric limit in Eq.\,\eqref{eps=}. However the LSS neutrino mass matrix does give successful predictions for all observables close to this limit, namely with ${m_\atm}/{m_\sol}\approx 10$ as compared to the $\mu\tau$ symmetric limit of ${m_\atm}/{m_\sol}=11$.

\section{Accidental implementations of $\mu\tau$ symmetry}
\label{sec:accidental}

The observation that $\mu\tau$-R symmetric $H_\nu$ is always real in a CP basis
---a basis where the generalized CP transformation \eqref{mutau-r:nuL} becomes canonical---
suggests an interesting way to obtain $\mu\tau$-U mixing accidentally\,\cite{rodejohann};
see appendix \ref{sec:cp.basis} for more details.
In other words, $\mu\tau$-U mixing may be obtained without implementing $\mu\tau$-R symmetry at the field level and hence allowing for nontrivial Majorana phases.
The key idea is that in a CP basis the automatic
$\mathtt{L}_\mu-\mathrm{L}_\tau$ symmetry of the charged leptons\,\cite{grimus:automatic},
\eq{
\label{L:mu-tau}
e_L\to e_L\,,\quad
\mu_L\to e^{i\theta}\mu_L\,,\quad
\tau_L\to e^{-i\theta}\tau_L\,,
}
is translated to a \emph{real orthogonal} symmetry $T$ (depending on $\theta$); see appendix \ref{sec:cp.basis}.
Therefore, in a CP basis, (i) there is a \textit{real} symmetry element $T$ that leaves the (lefthand) charged lepton mass matrix invariant and in such a basis (ii) the neutrino mass matrix $\bar{M}_\nu$ is diagonalizable by a \textit{real} orthogonal matrix.
This last property follows from the fact that in a CP basis, the decomposition \eqref{mutau-r:decomp} can be rewritten as 
\eq{
\label{A+iB}
\bM_\nu=\bar{A}+i\bar{B}\,,
}
with $\bar{A},\bar{B}$ being \emph{real} matrices.
The modified commutation relation \eqref{mod.comm} becomes simply the commutation relation $[\bar{A},\bar{B}]=0$; see appendix \ref{sec:cp.basis}.
Then $\mu\tau$-R symmetry in this basis simply corresponds to usual CP conservation and $\bar{B}=0$.

For the accidental implementation of $\mu\tau$-U proposed in Refs.\cite{rodejohann,patel}, one must promote the symmetry $T$ (one fixed and nontrivial $\theta$ is enough) to an actual flavor symmetry of the charged lepton sector.
One simple example for $T$ that is commonly found in nonabelian discrete groups such as $A_4$ or $S_4$ is
\eq{
\label{T:Z3}
T=\mtrx{0&0&1\cr 1&0&0 \cr 0&1&0}
\,.
}
At the same time, the neutrino mass matrix $\bar{M}_\nu$ in the same basis should be diagonalizable by a real matrix in an accidental way.
This can be arranged by a structure such as\,\cite{rodejohann}
\eq{
\bar{M}_\nu\sim \id +e^{i\alpha}(\text{real matrix}).
}
This would lead (for $\alpha\neq 0,\pi$) to nontrivial Majorana phases that cannot be obtained from $\mu\tau$-R symmetry on $M_\nu$.
Another option would be to also assume that $\bar{M}_\nu$ is invariant by one or two \emph{real} symmetries of a larger nonabelian symmetry such as $A_5$\,\cite{patel}.

We can go beyond the previous example by using the structure of vev alignments in indirect models\,\cite{indirect:cp} to achieve the structure in \eqref{A+iB}.
In a CP basis, we can write for the neutrino mass matrix,
\eq{
\label{bar-M:indirect}
\bar{M}_\nu=e^{i\eta}m_a u_1u_1^\tp +m_bu_2u_2^\tp + m_cu_3u_3^\tp\,,
}
where $u_i$, $i=1,2,3$ are \emph{real} vev alignments obtained from flavon vevs, $m_{a,b,c}$ are real parameters and $e^{i\eta}$ is a phase that can originate from spontaneous CP violation\,\cite{King:2015dvf}.
If we require the orthogonality condition
\eq{
\label{ortho}
u_1\perp u_2,u_3\,,
}
which is easy to be imposed in indirect models,
we clearly obtain two real commuting matrices in the form of
\eqali{
\bar{A}&=\cos(\eta) m_a u_1u_1^\tp +m_bu_2u_2^\tp+ m_cu_3u_3^\tp
\cr
\bar{B}&=\sin(\eta)m_au_1u_1^\tp\,,
}
which clearly obeys \eqref{A+iB}.
And then we achieved our goal of obtaining a $M_\nu$ with $\mu\tau$-U mixing without $\mu\tau$-R. 
Below in Subsec.\,\ref{sec:indirect} we give an explicit example compatible with current data.

\subsection{Example}
\label{sec:indirect}

Here we show an explicit example of the idea of accidental implementation in indirect models.
We use the CP basis where the mass matrix for the charged leptons comes from the vev alignment $(1,1,1)$ and then $M_l^\dag M_l$ is invariant by $T$ in \eqref{T:Z3} inside, e.g., $A_4$.
We then take the form \eqref{bar-M:indirect} with nontrivial $\eta$ and
\eq{
\label{u:indirect}
u_1=(0,1,1)^\tp
\,,\quad
u_2=(1,-1,1)^\tp
\,,\quad
u_3=(2,-1,1)^\tp
\,,
}
so that orthogonality \eqref{ortho} is ensured.
These vev alignments can be easily obtained for example from $A_4$ symmetry in the CSD framework\,\cite{Bjorkeroth:2014vha} in the real triplet basis of $A_4$.
For
\eq{
m_a\approx 6\,\text{meV}\,,\quad
m_b\approx 34\,\text{meV}\,,\quad
m_c\approx -11\,\text{meV}\,,
}
we obtain NO spectrum with observables within 3$\sigma$ and lightest mass $m_1\approx 
12\,\text{meV}$.
The phase $\eta$ can take any value and a nonzero value leads to nontrivial Majorana phases without disrupting the predictions of $\mu\tau$-U mixing.
In additional, the mixing obeys the TM1 form.

If required, we can go to the flavor basis in which the neutrino mass matrix is
\eq{
M_\nu=U_\om^\dag \bM_\nu U_\om^*\,,
}
where
\eq{
\label{Uw}
U_\om\equiv\frac{1}{\sqrt{3}}\mtrx{1&1&1\cr 1&\om&\om^2\cr 1&\om^2&\om}\,.
}
One can check that when $\eta$ is nontrivial, the mass matrix $M_\nu$ is not $\mu\tau$-R symmetric but $H_\nu$ is and obey the invariant conditions in \eqref{mutau:rephinv}.


\section{Conclusion}
\label{conclusions}

Motivated by the latest data which is consistent with maximal atmospheric mixing and maximal CP violation,
we have provided a timely survey of various results in $\mu\tau$ symmetry,
including several new observations and clarifications.
We have then applied the new results to the neutrino mass matrix associated with the Littlest Seesaw model,
and shown that it approximately satisfies a general form of $\mu\tau$ symmetry,
and that this is responsible for its approximate predictions of 
maximal atmospheric mixing and maximal CP violation in the lepton sector.

It is worth highlighting the new results and clarifications contained in this paper:
\begin{itemize}
\item We have carefully defined and distinguished between different kinds of $\mu\tau$ symmetry, namely 
$\mu\tau$ universal ($\mu\tau$-U) and $\mu\tau$ reflection ($\mu\tau$-R) symmetry, as applied to the PMNS matrix $V$,
the neutrino mass matrix $M_\nu$, and its  hermitean square $H_\nu\equiv M_\nu^\dag M_\nu$
(this is a clarification of known results).

\item We have shown that $\mu\tau$-R can be badly violated for the Majorana mass matrix $M_\nu$
without having large breaking of $\mu\tau$-U PMNS mixing,
and highlighted the role of Majorana phases (a clarification).

\item We have provided basis invariant conditions on $H_\nu$ leading to maximal atmospheric mixing and maximal CP violation (this is a clarification of the Harrison and Scott result).

\item We have presented a general form for $M_\nu$ leading to maximal atmospheric mixing and maximal CP violation
(a new result).

\item We have related pairs of $M_\nu$ and $\widetilde{M_{\nu}}$
related by $\mu\tau$ conjugation which have closely related predictions
(a new result).

\item We have applied these results to the Littlest Seesaw model and show that there is an approximate accidental $\mu\tau$ symmetry at work in this model (a new observation). 

\item Finally we have investigated new classes of models based on implementing $\mu\tau$-U without $\mu\tau$-R on 
the Majorana mass matrix $M_\nu$, based on the indirect approach and orthogonality on some vev alignments,
where CP is spontaneously broken as in the littlest seesaw models (a new direction).

\end{itemize}

In conclusion, if the present indications of maximal atmospheric mixing and maximal CP violation continue to 
hold up in future high precision neutrino experiments, then this may point to some exact or approximate
$\mu\tau$ symmetry underpinning the origin of neutrino mass and lepton mixing. We have reviewed this possibility,
making several new observations and clarifications, and applied our results to the Littlest Seesaw model.

\subsection*{Acknowledgements}
S.\,F.\,K.\ acknowledges the STFC Consolidated Grant ST/L000296/1 and the European Union's Horizon 2020 Research and Innovation programme under Marie Sk\l{}odowska-Curie grant agreements Elusives ITN No.\ 674896 and InvisiblesPlus RISE No.\ 690575.
C.\,C.\,N.\ acknowledges partial support by brazilian Fapesp through grant 2014/19164-6 and
CNPq through grants 308578/2016-3 and 454146/2017-5.

\appendix
\section{Proof that $\mu\tau$-R symmetric $H_\nu$ implies and is implied by $\mu\tau$-U PMNS mixing}
\label{proof1}

In this Appendix we review the connection between $\mu\tau$-U PMNS mixing and $\mu\tau$-R symmetric $H_\nu$ for both Dirac and Majorana neutrinos.

Explicitly,  for a generic neutrino mass matrix $M_\nu$ in the flavor basis to be diagonalizable by a $\mu\tau$-U mixing matrix satisfying \eqref{Vej} and \eqref{mutau-U}, 
it is \textit{necessary} and \textit{sufficient} that $H_\nu=M_\nu^\dag M_\nu$ is 
complex and additionally invariant by $\mu\tau$-reflection symmetry:%
\subeqali[Mnu2:mutau-r]{
\label{Mnu2:mutau-r:1}
(H_\nu)^*&\neq H_\nu\,
\\
\label{Mnu2:mutau-r:2}
P_{\mu\tau}^\dag H_\nu P_{\mu\tau}&=(H_\nu)^*
\,,
}
after allowing for appropriate \textit{rephasing transformations}.
The second condition is just \eqref{mutau-r:A}.
The first condition is important to distinguish $\mu\tau$-R symmetry from the simpler $\mu\tau$ interchange symmetry, which predicts $\theta_{13}=0$ and is now excluded.

Necessity is straightforward to check: once the PMNS mixing is $\mu\tau$-U, apply rephasing transformations to write it in the form \eqref{U0}.
In that phase convention,
the diagonalization relation $H_\nu =V_0\diag(m_i^2)V_0^\dag$ allows us to check 
\eqref{Mnu2:mutau-r:2} because $V_0$ in \eqref{U0} enjoys the property
\eq{
\label{Pmutau.U0}
P_{\mu\tau}^\dag V_0=V_0^*\,.
}
The complexity of $H_\nu$ in \eqref{Mnu2:mutau-r:1} follows because $V_0$ is essentially complex.

Now, sufficiency is also easy to prove\,\cite{mutau-r:HS}.
However, we detail the proof to emphasize where condition \eqref{Mnu2:mutau-r:1} plays a 
role.
We adapt the proof of Grimus and Lavoura \cite{mutau-r:GL} from $M_\nu$ to $H_\nu$.
The symmetry \eqref{Mnu2:mutau-r:2} implies that if $u_i$ is an eigenvector of $H_\nu $ 
with eigenvalue $m_i^2$, then $P_{\mu\tau}^\tp u_i^*$ is also an eigenvector with the same eigenvalue.
Then it follows that $P_{\mu\tau}^\tp u_i^*=u_i$, assuming nondegenerate masses and conveniently 
choosing the first component of $u_i$ to be real and positive.
The previous relation is equivalent to \eqref{Pmutau.U0} which is consistent with 
\eqref{mutau-U}.
The last step is to check if condition \eqref{Vej} is guaranteed.
Let us use the basis where the $(e\mu)$ and $(e\tau)$ entries of $H_\nu $ are real and 
positive so that the $(\mu\tau)$ entry is the only complex entry, following from 
\eqref{Mnu2:mutau-r:1}.
Assume in addition that the first component of some $u_i$ is zero.
Then the eigenvector relation $(H_\nu u_i)_1=m_i^2(u_i)_1=0$ together with the form
\eqref{Mnu2:mutau-r:form} implies that $u_i$ should be of the form $u_i\sim 
(0,1,-1)^\tp$.
Checking the second and third components of the eigenvector relation for $u_i$, we 
conclude that $(H_\nu )_{\mu\tau}$ should be real, contradicting the hypothesis.

\section{CP basis}
\label{sec:cp.basis}

When $H_\nu$ or $M_\nu$ are $\mu\tau$-R symmetric, as in Eqs.\,\eqref{Mnu2:mutau-r:form} or \eqref{Mnu:mutau-r:form}, it is always possible to change basis so that it becomes real and the
$\mathtt{L}_\mu-\mathtt{L}_\tau$ symmetry in \eqref{L:mu-tau} becomes a real symmetry;
see appendix B of Ref.\cite{cp.mutau}.
Such a basis is also a CP basis (or real basis) with respect to the generalized CP symmetry in \eqref{mutau-r:nuL} where the latter becomes the canonical CP transformation\,\cite{gunion.haber}.
In such a basis, the part of the Lagrangian invariant by \eqref{mutau-r:nuL} will only contain real parameters. See, however, caveat in \cite{cp4}.

One possible basis transformation that takes us to one CP basis is given by the matrix 
\eq{
\label{Umutau}
U_{\mu\tau}\equiv
\left(
\begin{array}{ccc}
 1 & 0 & 0 \\
 0 & \frac{1}{\sqrt{2}} & -\frac{i}{\sqrt{2}} \\
 0 & \frac{1}{\sqrt{2}} & \frac{i}{\sqrt{2}} \\
\end{array}
\right)
\,,
}
as, in the new basis, the $\mu\tau$-R involves the trivial interfamily transformation
\eq{
U_{\mu\tau}^\dag P_{\mu\tau}U_{\mu\tau}^*=\id_3\,.
}

If we perform the explicit basis transformation \eqref{Umutau} to $H_\nu$ of the form \eqref{Mnu2:mutau-r:form}, we obtain
\eq{
\label{bHnu}
\bH_\nu=U_{\mu\tau}^\dag H_\nu U_{\mu\tau}
=
\left(
\begin{array}{ccc}
 A & \sqrt{2} \re(D) & \sqrt{2} \im(D) \\
 \sqrt{2} \re(D) & B+\re(C) & \im(C) \\
 \sqrt{2} \im(D) & \im(C) & B-\re(C) \\
\end{array}
\right)\,.
}
By inspection we see that the resulting matrix $\bH_\nu$ is a \emph{generic} real 
and symmetric matrix, except that it is required to be a positive semidefinite 
matrix due to its nature as a hermitean square.
Analogously, $M_\nu$ of the form \eqref{Mnu:mutau-r:form} is transformed to 
\eqref{bHnu} with uppercase letters replaced by lowercase letters.
Then components $A,B$
in the general form \eqref{mutau-r:decomp} become two \emph{generic} real matrices 
$\bar{A},\bar{B}$ and the modified commutation condition \eqref{mod.comm} becomes 
simply the commutation relation $[\bar{A},\bar{B}]=0$.

One can also check explicitly that the $\mathtt{L}_\mu-\mathtt{L}_\tau$ symmetry in \eqref{L:mu-tau} is transformed to
\eq{
\label{real.T}
T=U_{\mu\tau}^\dag\mtrx{1&&\cr &e^{i\theta}&\cr &&e^{-i\theta}}U_{\mu\tau}
=\mtrx{1&&\cr&\cos\theta&\sin\theta\cr &-\sin\theta&\cos\theta}
\,,
}
in the CP basis connected by \eqref{Umutau}.
The transformation \eqref{real.T} is a real and orthogonal matrix.
Any further real basis change will keep $T$ as a real orthogonal matrix.

Clearly, there are infinitely many CP bases reachable by any further \textit{real} 
basis transformation from the basis defined by \eqref{Umutau}.
This class exhausts all possible CP bases, except for additional discrete symmetries of the theory, which we can factor.
The transformation \eqref{Uw} is another common example of this class.


\end{document}